\documentclass[%
 aip,
 amsmath,amssymb,
 reprint,%
]{revtex4-1}

\usepackage{graphicx}
\usepackage{dcolumn}
\usepackage{bm}

\usepackage[utf8]{inputenc}
\usepackage[T1]{fontenc}
\usepackage{mathptmx}

\begin{document}

\preprint{AIP/123-QED}

\title{6 Gbps real-time optical quantum random number generator based on vacuum fluctuation}

\author{Ziyong Zheng}

\author{Yichen Zhang}%
\email{zhangyc@bupt.edu.cn.}

\author{Weinan Huang}

\author{Song Yu}
\email{yusong@bupt.edu.cn.}
\affiliation{State Key Laboratory of Information Photonics and Optical Communications, Beijing University of Posts and Telecommunications, Beijing, 100876, China}

\author{Hong Guo}
\affiliation{State Key Laboratory of Advanced Optical Communication Systems and Networks, School of Electronics Engineering and Computer Science, and Center for Quantum Information Technology, Peking University, Beijing, 100871, China}%

\date{\today}

\begin{abstract}
We demonstrate a 6 Gbps real-time optical quantum random number generator by measuring vacuum fluctuation. To address the common problem that speed gap exists between fast randomness generation and slow randomness extraction in most high-speed real-time quantum random number generator systems, we present an optimized extraction algorithm based on parallel implementation of Toeplitz hashing to reduce the influence of classical noise due to the imperfection of devices. Notably, the real-time rate of randomness extraction we have achieved reaches the highest speed of 12 Gbps by occupying less computing resources and the algorithm has the ability to support hundreds of Gbps randomness extraction. By assuming that the eavesdropper with complete knowledge of the classical noise, our generator has a randomness generation speed of 6.83 Gbps and this supports the generation of 6 Gbps information-theoretically provable quantum random numbers, which are output in real-time through peripheral component interconnect express interface.
\end{abstract}

\maketitle

\section*{Introduction}

	Random numbers are the basis for applications in statistics, simulation \cite{Ferrenberg1992}, cryptography \cite{Gennaro1621063} and fundamental science \cite{Brunner2014}. The randomness of random numbers will directly affect the overall security of corresponding application systems. Classical random number generation methods, for example pseudo random number generators (RNG) based on determined algorithms \cite{Nisan1988Hardness}, provide a cost-efficient and portable method to produce pseudo random numbers at a high speed, which satisfies the demand for random numbers of most applications. However, due to the deterministic and predictable features of the algorithms, pseudo RNG are not suitable for certain applications where true randomness is required. In cryptography applications, random numbers with untrusted randomness will result in safety issues, since hackers can access the information of random numbers and thus crack the encryption systems \cite{Bouda2012Weak}. The rapid development of quantum cryptography technologies such as quantum key distribution \cite{Weedbrook2012Gaussian,Scarani2012The,Diamanti2016Practical,Zhang2017Continuous,Gisin2002quantum} which requires secure, real-time and high-speed random number generation, unarguably accelerate the researches about true random number generation.

	Distinct from pseudo RNG, the optical quantum random number generators (QRNG) based on the intrinsic randomness of fundamental quantum processes are guaranteed to produce nondeterministic and unpredictable random numbers \cite{Ma2016Quantum,Bera2017Randomness,Herrero2017quantum}. Such advantages attract researchers' attention and many related generator protocols are proposed. Substantial practical QRNG protocols have been demonstrated to realize high-speed random number generation with relatively low cost, including measuring photon path \cite{Jennewein2000A,Andr2000Optical}, photon arrival time \cite{Michael2009Photon, Nie2014Practical, Dynes2008A, Wahl2011An,  Ma2005Random}, photon number distribution \cite{Wei2009Bias,Furst2010High,Applegate2015Efficient,Ren2011Quantum,Yan2014Multi}, vacuum fluctuation \cite{Gabriel2010A,Shen2010Practical,Symul2011Real,Haw2015Maximization,Bingjie2017High,Santamato2017An,Zhou2017Practical}, phase noise \cite{Qi2010Highspeed,Guo2010Truly,Xu2012Ultrafast,Abellan2014Ultra,Nie2015The,Yang2016A,Zhang2016Note,Liu2017117} and amplified spontaneous emission noise of quantum states\cite{Williams2010Fast,Li2011Scalable,Martin2015Quantum,Liu2013Implementation,Wei2012High,Qi2017True}, etc. Typically, protocol based on the measurement of vacuum fluctuation is a more applied and valuable QRNG protocol, for its convenience of state preparation, insensitivity of detection efficiency and high generation speed.

	QRNGs based on measuring vacuum fluctuation are generally realized by applying homodyne scheme \cite{Gabriel2010A,Shen2010Practical,Symul2011Real,Haw2015Maximization,Bingjie2017High,Santamato2017An,Zhou2017Practical}, which realize the measurement of the quadrature amplitude of vacuum state so as to generate true random numbers. Their security is ultimately guaranteed by the laws of quantum physics and can be achieved by applying ideal devices. But actually the practical devices equipment in the QRNG systems can not always meet the requirements, which will inevitably introduce classical noise, and ultimately lead to system security reduction.
	
	The universal hash function families including Toeplitz hashing \cite{Impagliazzo1989Pseudo,Mansour1990The} have proved to be information-theoretically secure, and they are widely used to eliminate the influence of classical noise that compromise the security of generated random numbers\cite{Ma2012Postprocessing}. The postprocessing process is generally called randomness extraction. Toeplitz hashing matrix is often chosen as a randomness extraction algorithm because of its low computation and implementation complexity. Up to now, there are many Toeplitz hashing implementations realized on different platforms including central processing unit (CPU) devices, general processor unit (GPU) devices and field programmable gate array (FPGA) devices, etc. The implementations on CPU devices transform the Toeplitz hashing algorithm to fast Fourier transform (FFT) algorithm while their speeds are limited to relatively small values, such as 441 kbps\cite{Ma2012Postprocessing} and 1.6 Mbps\cite{Nie2015The}. The GPU implementation realizes the parallel computation of many FFT threads in a GPU device and thus achieves a real-time extraction rate of 1.35 Gbps \cite{Wang2018High}. However, these speeds still do not match the practical quantum key distribution system's demand for high-speed real-time random number generation.

	Considering the advantages of FPGA with parallel computing, the Toeplitz hashing implementation in FPGA devices may be a most promising method to achieve fast randomness extraction speed so as to support high-speed real-time random number generation. The work in Ref.\cite{Zhang2016Note} introduces a QRNG based on the measurement of laser phase fluctuations and realizes a 3.36 Gbps real-time randomness extraction and finally supports a 3.2 Gbps real-time random number generationdue to the rate limitation of	its small form-factor pluggable interface. Its randomness extraction algorithm applied in this work transforms the whole Toeplitz matrix into submatrices and performs multiplication operation between the submatrices and the input raw data, which improves the performance of randomness extraction. While multiplication operation between the submatrices and the input raw data will still consume certain computing resources, which will significantly limit the final randomness extraction speed on a given device where computing resources are finite. To address this issue, an optimized Toeplitz hashing algorithm is required to reduce computing resource consumption and finally supports a faster extraction speed.
	
	In this paper, we present a 6 Gbps real-time QRNG based on measuring vacuum fluctuation using the homodyne scheme. To fill the gap between the rapid randomness generation and the slow randomness extraction, we realize an optimized Toeplitz hashing algorithm to support the realization of high-speed generators. The real-time randomness extraction speed we have realized reaches 12 Gbps by occupying less computing resources and the algorithm has the ability to support hundreds of Gbps randomness extraction, which is faster than the ever reported 3.36 Gbps randomness extraction method \cite{Zhang2016Note}. Assuming that the eavesdropper with complete knowledge of the classical noise, our generator has a randomness generation speed of 6.83 Gbps and this supports the generation of 6 Gbps information-theoretically provable quantum random numbers, which are output in real-time through the peripheral component interconnect express (PCIE) interface with a security parameter of ${\rm{1}}{\rm{.03}}{{\rm{e}}^{ - 128}}$.

	\begin{figure}[t]
		\centering
		\includegraphics[width = 7 cm]{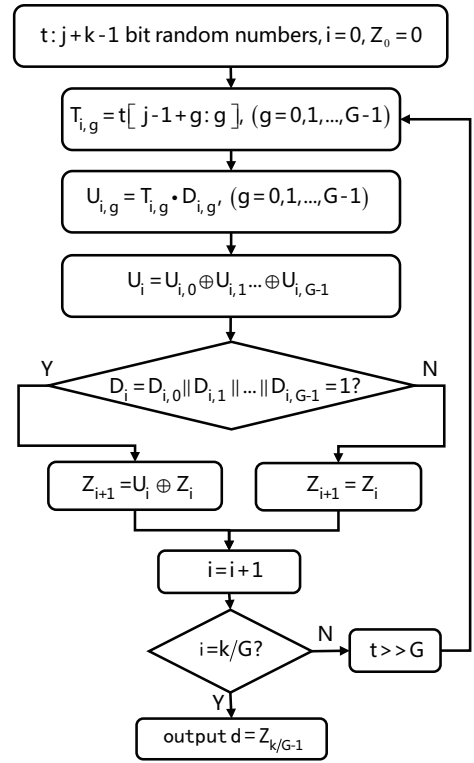}
		\caption{\label{algorithm_chart}Real-time randomness extraction algorithm of each calculation block.}
	\end{figure}

\section*{Real-time Randomness Extraction Algorithm}

	For practical QRNG system, the electrical noise existing in the actual system will affect the security of the raw data, so a corresponding random number extraction operation is very important to eliminate the influence of the classical noise. The random extraction operation is usually carried out on the basis of the min-entropy estimation which helps to characterize the extractable randomness and the true random numbers will be generated after randomness extraction. For a real-time system, the extraction rate is usually the bottleneck of the whole system. Therefore, it is very important to design a high-speed random extraction algorithm so as to improve the overall performance of the system.
	
	As one of the universal hash functions, Toeplitz hashing matrix is often chosen as an extraction algorithm because of its low computation and implementation complexity. A binary Toeplitz matrix $T$ with a size of $j \times k$ can be constructed by $j + k - 1$ random bits for the reason that each descending diagonal of Toeplitz matrix is the same. The collision probability of such a Toeplitz matrix, which indicates the probability of having the same output for different input raw data, is determined by the number of rows and the number of columns in the Toeplitz matrix. The collision probability of a certain Toeplitz hash is equal to $k \cdot {2^{( - j + 1)}}$, where the number of columns $k$ equals to the input data length and the number of rows $j$ indicates the output data length. So a relatively small collision probability can be obtained by selecting suitable $k$ and $j$ values.
	
	\begin{figure*}[t] 
		\centering
		\includegraphics[width = 16 cm]{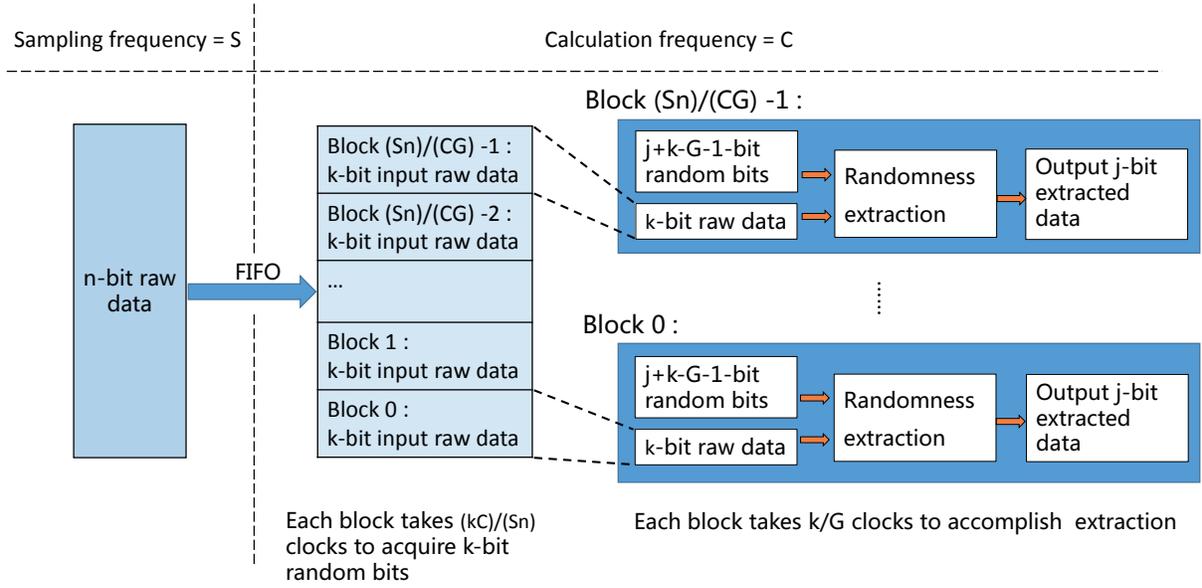}
		\caption{\label{multi_module_scheme} The process of randomness extraction algorithm that multi blocks work in parallel. To bridge the speed gap, the sampled high speed $n$-bit raw data is transferred into multiple low speed preprocessed $k$ bit data blocks through one or several memory queue modules called First Input First output (FIFO). $(Sn)/(C G)$ calculation blocks based on the algorithm introduced in FIG.$\ref{algorithm_chart}$ are cyclically called from block 0 to block  $(Sn)/(CG) - 1$ as soon as each block finishes raw data acquiring. The reason that the random extraction works at a frequency of $C$ rather than $S$ is that the general FPGA development board does not support too fast clock.}
	\end{figure*}
				
	The parallel computing advantages of FPGA make it widely used in high-speed computing applications. However, it is difficult for FPGA to perform large-scale matrix computing directly due to the limitation of FPGA resources. The real-time randomness extraction algorithm applied in Ref.\cite{Zhang2016Note} transforms the whole Toeplitz matrix into submatrices and performs multiplication operation between the submatrices and the input raw data. While multiplication operation between the submatrices and the input raw data will still consume certain computing resources when the submatrices are large. Considering the advantages of FPGA with parallel computing character, we optimize the Toeplitz hashing algorithm by transforming the multiplication between submatrices and the input raw data into exclusive or (XOR) operation between the columns of the Toeplitz hashing matrix, which can significantly reduce the resource consumption and support higher extraction rate. The multiplication operation of Toeplitz matrix $T$ and raw data $D$ can be transformed into the XOR operation between columns of $T$. 
		
	 As shown in FIG.\ref{algorithm_chart}, $j+k-1$ random bits are utilized as the random number seeds. Each clock $G$ columns of the Toeplitz matrix can be  separately represented as ${T_{i,g}} = t\left[ {j - 1 + g:g} \right],g = 0,1,2,...,G-1$. The $G$ raw data bits ${D_{i, g}} =1$ or 0 will simultaneously lead to the corresponding intermediate variable ${U_{i,g}}$ = ${T_{i,g}}$ or ${U_{i, g}}$ = 0, and ${U_i} = {U_{i, 0}}{\kern 1pt}  \oplus {U_{i, 1}}... \oplus {U_{i, G-1}}$ will be achieved. If ${D_i} = {D_{i, 0}}{\kern 1pt} \parallel {D_{i, 1}}\parallel ...\parallel {D_{i, G-1}} = 1$, another intermediate variable ${Z_{i + 1}}{\kern 1pt} {\kern 1pt} {\kern 1pt}  = {\kern 1pt} {\kern 1pt} {U_i}{\kern 1pt} {\kern 1pt} {\kern 1pt} {\kern 1pt}  \oplus {\kern 1pt} {\kern 1pt} {\kern 1pt} {\kern 1pt} {Z_i}$. Otherwise, if ${D_i} = {D_{i, 0}}{\kern 1pt} \parallel {D_{i, 1}}\parallel ...\parallel {D_{i, G-1}} = 0$, then ${Z_{i + 1}}{\kern 1pt} {\kern 1pt} {\kern 1pt}  = {\kern 1pt} {\kern 1pt} {\kern 1pt} {\kern 1pt} {Z_i}$. The value of counter $i$ increases by 1 per round. If $i$ equals to $k/G$, the calculated result $d{\kern 1pt} {\kern 1pt} {\kern 1pt}  = {\kern 1pt} {\kern 1pt} {\kern 1pt} {Z_{k/G-1}}$ outputs. Otherwise, $t$ will shift $G$ bits to the right and restart the assignment of ${T_{i,g}}$.  The output $d$ is the final extracted $j$ bit random numbers. Notably, the column number for each XOR operation can be not only $G$, but also the integer multiple of $G$. Usually, we set the value of $G$ equal to the integral rate of the sampling precision $n$ to simplify the calculation process. Assuming that the calculation frequency of the extractor is $C$, such a random extraction module can realize real-time processing of raw data with $CG$ rate.

	\begin{figure*}[t]
		\centering
		\includegraphics[width = 16 cm]{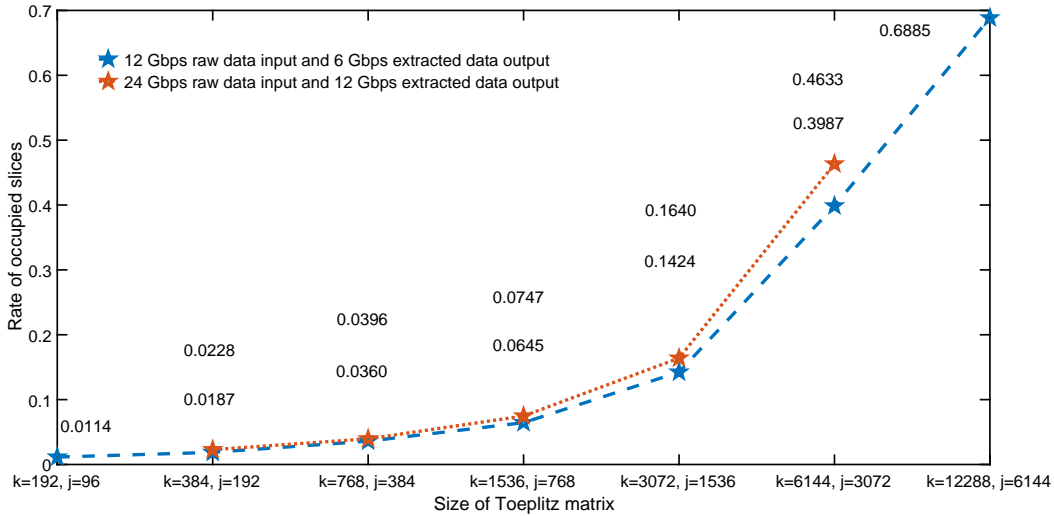}
		\caption{\label{Computation_resource}Computing resources consumption test result of the algorithm. We focus on the resource of slice that constitutes the basic logic unit of FPGA. The test is performed on the Xilinx KC705 evaluation board, which has 50950 slices. $k$ equals to the input data length and the column number of matrix $T$. $j$ indicates the output data length together with the row number of matrix $T$. The blue line is the test result that using different Toeplitz Matrix to deal with 12 Gbps input raw data. The input data is divided into 8 blocks to be calculated separately. The red line shows the results of dealing with 24 Gbps input raw data, which is divided into 16 blocks to be calculated separately.}
		
	\end{figure*}
			
	However, the FPGA platform supports calculation speed $C$ that is much slower than the sampling frequency $S$. So it is necessary to transform the sampled high speed $n$ bit raw data into multiple low speed preprocessed $k$ bit data blocks through a caching medium called First Input First Output (FIFO). 
	
	As shown in FIG.\ref{multi_module_scheme}, these blocks are numbered sequentially as 0, 1, 2, …, $(Sn)/(CG)-1$. The transformation progress can be divided into two steps. Firstly, we use FIFO to convert $n$ bit sampled data at a frequency of $S$ to ${{Sn/C}}$ bit data at a frequency of $C$. Due to the limitation of FPGA, the value of $S/C$ is usually set as $2^i$ and the value of $i$ can be set as -3, -2, -1, 0, 1, 2, or 3. If the value of $S/C$ needs to be greater than 8, more FIFOs will be cascaded to implement such a function. After the first step, the bit rate on the output side of the FIFO is $Sn$, which is much faster than randomness extraction speed, $CG$, of each block. So secondly, we will distribute the $(Sn)/C$ bit data from the output port of the FIFO to $(Sn)/(CG)$ independent calculation blocks sequentially by controlling the output enable signal. Thus each block will take $(kC)/(Sn)$ clocks to receive $k$ bit data sequentially. We should notice that the frequency of ${{Sn/C}}$ bit data that each block received is $C$ instead of $S$. The following processing will also work at a frequency of $C$.
	
	After one block finishes storing $k$ bit data, this block will stop data storage and the next block begins to store the next $k$ bit data. When the last block, block $(Sn)/(CG)-1$, finishes data storage, the first block, block $0$, begins data storage again, which constitutes a loop of data storage. It will take $k/G$ clocks for all $(Sn)/(CG)$ blocks to realize data storage sequentially.
	
	Finally, we will apply the randomness extraction algorithm introduced in FIG.\ref{algorithm_chart} in each block. Each clock $G$ bit raw data will be used to control the extraction progress, so that each block with $k$ bit input raw data will take $k/G$ clocks to accomplish the randomness extraction and output $j$ bit extracted random numbers. The extraction operation of each block begins as soon as the block finishes data storage and they are calculated independently. Usually the value of $G$ can be set as the integer multiple of the sampling precision $n$, and in our setup this value is equal to $n$ to simplify the calculation process.
			
	Computing resource consumption test of the algorithm is performed on the Xilinx KC705 evaluation board to find a suitable parameter configuration scheme. It is easy to understand that the extraction speed of the extractor is related to the number of blocks in parallel operation together with the hardware resource of the computing platform. The speed of randomness extraction increases linearly as the number of blocks increases under a given calculation frequency, while the occupied computing resources will also increase. The results of the test are shown in FIG.\ref{Computation_resource}. We set ${j \mathord{\left/ 	{\vphantom {j k}} \right.
	\kern-\nulldelimiterspace} k}= 1/2 $ and $G=n=12$ to test how different $k$ and $j$ values affect the occupancy of computing resources. For a specified block number and calculating frequency $C$, the greater the value of $k$ and $j$, the more computing resources will be consumed and the increase of $k$ and $j$ values is helpless to the calculation speed. While greater $k$ and $j$ values help to gain a smaller security parameter when the extractable randomness value is determined, which will be analyzed in detail in the next section. Notably, if we use Toeplitz matrix with a small size, i.e., $192 \times 384$, it is possible to realize real-time processing of hundreds of Gbps raw data with this algorithm. In this experiment, we ensure that the calculated security parameter is small enough by selecting appropriate $k$ and $j$, while at the same time make smaller use of computing resources. The randomness extractors supporting $24$ Gbps and 12 Gbps real-time input raw data are realized and their slice consumption results are shown in FIG.\ref{Computation_resource}, where slice constitutes the basic logic unit of FPGA. Taking one of the extractors for an example, it will occupy 16.40\% slices when the size of Toeplitz hashing matrix is $1536 \times 3072$ and 16 blocks are called to realize the real-time randomness extraction of 24 Gbps real-time input raw data, which means our algorithm is very promising to support even faster randomness extraction in our platform i.e., KC705 evaluation board.

\section*{Experimental Implementation}
		\begin{figure*}[t]
			\centering
			\includegraphics[width = 14 cm]{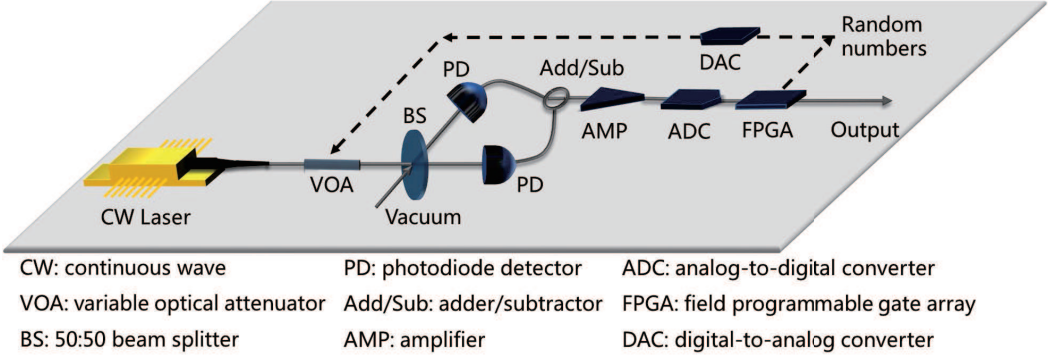}
			\caption{\label{complex_flow_chart}Experimental demonstration of real-time optical QRNG based on vacuum fluctuation. The CW beams emitted by the laser diode enter one input port of the 50:50 BS. The other input port of the BS is blocked to provide the vacuum state. A following measurement operation is realized by a homodyne detector and an ADC. The measurement result is finally processed by a randomness extractor to distill the final random bits.}
		\end{figure*}	
		\begin{figure}[t]
			\centering
			\includegraphics[width = 9 cm]{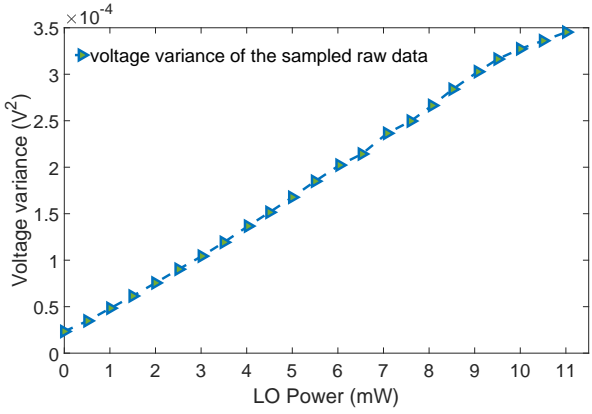}
			\caption{\label{variance_lopower}Variance vs LO power. This figure shows the voltage variance of the sampled raw data as a function of the LO power. The LO power is increased by adjusting the variable attenuator from 0 mW with a step size of 0.5 mW. The voltage variance of the raw data enhances linearly with LO power in the range of 0 mW to 9.5 mW. The slope of the trend curve will decrease when the LO power is larger than 9.5 mW.}
		\end{figure}
		\begin{figure}[t]
			\centering
			\includegraphics[width = 8 cm]{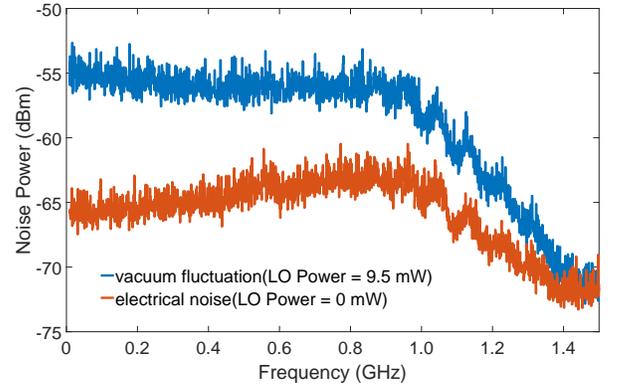}
			\caption{\label{sanli_to_ele3} The power spectrum density of the vacuum
				fluctuations when the LO power is 9.5 mW (blue line) and the electrical
				noise when the LO power is 0 mW (red line). The 3 dB bandwidth range ${{f_0}}$ is approximately 0-1 GHz. The corresponding result is acquired by using the spectrum analyzer to detect the output signal of the detector.}
		\end{figure} 	
		
	To realize a practical secure, real-time and high-speed QRNG, we implement an all-in-fiber setup with several off-the-shelves devices and a self-developed analog-to-digital converter (ADC) card, and apply the optimized real-time randomness extractor to generate true random numbers by measuring vacuum fluctuation. The block diagram is shown in FIG.\ref{complex_flow_chart}.

	A 1550-nm fiber-coupled laser (NKT Basic E15, linewidth 100 Hz) serves as the local oscillator (LO) and is connected to one input port of the 50:50 beam splitter (BS). While the other input port is blocked to provide the vacuum state. The two output ports of the beam splitter are optically coupled to the two input ports of a balanced homodyne detector (Thorlabs PDB480C, the measurement bandwidth limited to 1 GHz by low-pass filter). The measurement results of the balanced homodyne detector are finally sampled by a 12 bit ADC card (ADS5400, sampling frequency 1 GHz, sampling precision 12 bit and input voltage range $2R$ reduced from 2 VPP to 1.5 VPP by the pre-amplifier of ADC card) to acquire the raw data in real time. A following randomness extractor based on the optimized algorithm is used to perform extraction simultaneously with raw data acquiring.

	Assuming in the worst case that the adversary can listen to and control over the classical noise, Haw et al \cite{Haw2015Maximization} has derived the maximum min-entropy of discretized measurement signal ${M_{dis}}$ conditioned on classical noise $E$, which is
	\begin{equation}
		\label{eq1}
		\begin{split}
			{H_{\min }}({M_{dis}}|E) =  - {\log _2}[\max ({c_1},{c_2})],
		\end{split}
	\end{equation}
	where ${c_1} = \frac{1}{2}[erf(\frac{{{e_{\max }} - R + {{3\delta } \mathord{\left/
				{\vphantom {{3\delta } 2}} \right.
				\kern-\nulldelimiterspace} 2}}}{{\sqrt 2 {\sigma _Q}}}) + 1]$, ${c_2} = erf(\frac{\delta }{{2\sqrt 2 {\sigma _Q}}})$, and the value of $M$ is the superposition of quantum noise $Q$ and classical noise $E$. $R$ equals to half of the input voltage range of ADC card and $\delta=2R/(2^n)$, where $n$ is the sampling precision of ADC card. ${\sigma _Q}$ indicates the value of the standard deviation of quantum fluctuation. This result is based on the assumption that quantum randomness is independent of classical noise and the value of classical noise is within a finite interval, i.e., $ - 5{\sigma _E} \le e \le 5{\sigma _E}$, which is valid for 99.9999\%.

	We simplify the analysis progress and ensure that the system works under safety conditions that ${c_1} \le {c_2}$, in which the comparison between ${c_1}$ and ${c_2}$ will indicate whether the min-entropy evaluation utilizes the correct maximum guessing probability. In this case, we can simplify the conditional min-entropy as 
	
	\begin{equation}
	\label{eq2}
	\begin{split}
		{H_{\min }}({M_{dis}}|E) =  - {\log _2}[erf(\frac{\delta }{{2\sqrt 2 {\sigma _Q}}})], 
	\end{split}
	\end{equation}
	in which we find that ${H_{\min }}({M_{dis}}|E)$ increases with decreasing $\delta $ and increasing $\sigma _Q$. For a given QRNG system, its sampling precision $\delta $ is determined, thus we consider how the constructed components influence the value of $\sigma _Q$.

	\begin{figure*}[t]
			\centering
			\includegraphics[width = 16 cm]{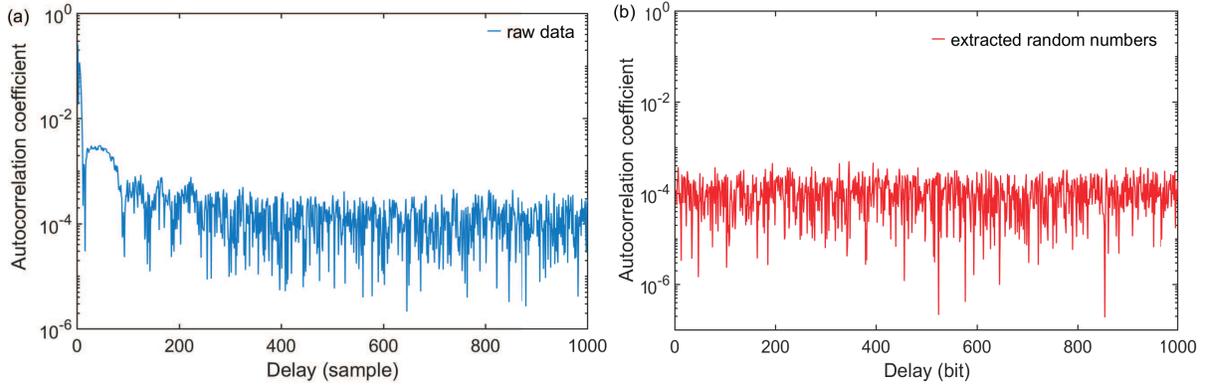}
			\caption{\label{auto_test} Autocorrelation of $10^7$ raw data (a) and $10^7$ bit extracted random numbers (b). The imperfect homodyne detector without steep sideband will inevitably compromise the autocorrelation performance, which leads to higher values of low-order autocorrelation. The autocorrelation existing in the raw data is well eliminated by the Toeplitz hashing extraction process.}
	\end{figure*}		
	\begin{figure}[t]
			\centering
			\includegraphics[width = 8 cm]{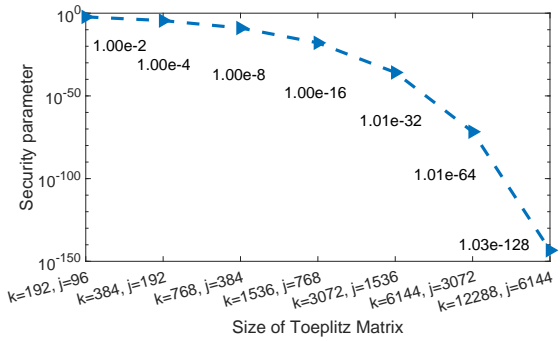}
			\caption{\label{security_bound} Calculated security parameter values for different Toeplitz matrix sizes when the extractable randomness of input raw data is $56.92\%$.}
	\end{figure}
			
	A suitable LO power improve the value of $\sigma _Q^2$ as Ref.\cite{Bingjie2017High} introduced. The LO power is increased by adjusting the variable attenuator from 0 mW with a step size of 0.5 mW to seek for the optimal $\sigma _Q^2$. Simultaneously the voltage variance of each measured raw data is calculated and recorded, as shown in FIG.\ref{variance_lopower}. When the LO power is set to 0 mW, the measured voltage variance of $10^8$ successive raw data is treated as $\sigma _E^2$, which has an average calculated value of ${\rm{3}}{\rm{.13}}{{\rm{e}}^{ - 5}}{{\rm{V}}^2}$. FIG.\ref{variance_lopower} indicates that the voltage variance of the raw data enhances linearly with LO power in the range of 0 mW to 9.5 mW. While the slope of the trend curve will decrease when the LO power is larger than 9.5 mW. An average  $\sigma _M^2$ value of the corresponding $10^8$ successive data can be obtained as ${\rm{3}}{\rm{.16}}{{\rm{e}}^{ - 4}}{{\rm{V}}^2}$ by setting the LO power to 9.5 mW. For our ADC with 12 bit sampling precision and 1.5 VPP input voltage range, its variance of quantization errors is calculated as ${({\delta  \mathord{\left/
				{\vphantom {\delta  {12}}} \right.
				\kern-\nulldelimiterspace} {12}})^2} = ({{1.5} \mathord{\left/
			{\vphantom {{1.5} {(4096}}} \right.
			\kern-\nulldelimiterspace} {(4096}} \cdot 12){)^2} = {\rm{9}}.{\rm{3132}}{{\rm{e}}^{{\rm{ - 10}}}}{{\rm{V}}^2}$. Quantization errors exist in the discretization of $M$ and $E$, so the secure upper bound of $\sigma _Q^2$ can be calculated as $\sigma _M^2-\sigma _E^2-2{({\delta  \mathord{\left/
				{\vphantom {\delta  {12}}} \right.
				\kern-\nulldelimiterspace} {12}})^2}={\rm{2}}{\rm{.8457}}{{\rm{e}}^{{\rm{ - 4}}}}{{\rm{V}}^2}$ when considering the influence of ADC's quantization errors. Substitute $\sigma _Q^2$ into Eq. (\ref{eq2}) and ${H_{\min }}\left( {M_{dis}|E}\right)$ is thus calculated as 6.83 bits per sample, which means that $56.92 \%$ random bits can be generated from each sample and the real-time quantum randomness generation speed reaches 6.83 Gbps. It is noteworthy that a self-designed 1.2 GHz balanced homodyne detector is used to build a quantum random number generator system and it has achieved a min-entropy value of 6.53 bits per sample\cite{ZHANG20181.2G}.

	The quantum noise is well above the classical noise at the output end of the homodyne detector when the LO power is set to 9.5 mW, as shown in FIG.\ref{sanli_to_ele3}. The following 12 bit ADC with a sampling frequency of 1 GHz then quantifies the signals into digital data. 8 extracting operation blocks with a synchronized calculation frequency $C$ of 125 MHz are performed in parallel on FPGA in real time. For our implementation of Toeplitz hashing extraction, we set $j = 6144$ and $k = 12288$ so that the extraction ratio is
	${j \mathord{\left/
			{\vphantom {j k}} \right.
			\kern-\nulldelimiterspace} k}{\rm{ = }}50\%  $, which is smaller than $56.92 \%$ derived from the calculated ${H_{\min }}\left( {M_{dis}|E}\right)$.
	
	Thus the information theoretic security parameter $\varepsilon$, which means the statistical distance between the extracted random sequence and the uniform sequence, can be calculated by the leftover hash lemma,
	$j = k \cdot {{{H_{\min }}\left( {M_{dis}|E} \right)} \mathord{\left/
			{\vphantom {{{H_{\min }}\left( {M_{dis}|E} \right)} n}} \right.
			\kern-\nulldelimiterspace} n} - 2 \cdot {\log _2}\left( {{\rm{ }}{1 \mathord{\left/
				{\vphantom {1 \varepsilon }} \right.
				\kern-\nulldelimiterspace} \varepsilon }} \right)$. The calculated $\varepsilon$ is approximately ${\rm{1}}{\rm{.03}}{{\rm{e}}^{ - 128}}$ in our setup. With such configuration, the real-time random number generation rate can finally achieve 6 Gbps. Notably, for different Toeplitz matrix sizes, the calculated security parameter values are quite different, as shown in FIG.\ref{security_bound}. Different Toeplitz matrices can be selected to perform extraction according to the actual available resources and different security requirements.
	
			\begin{figure}[t]
				\centering
				\includegraphics[width = 7.5 cm]{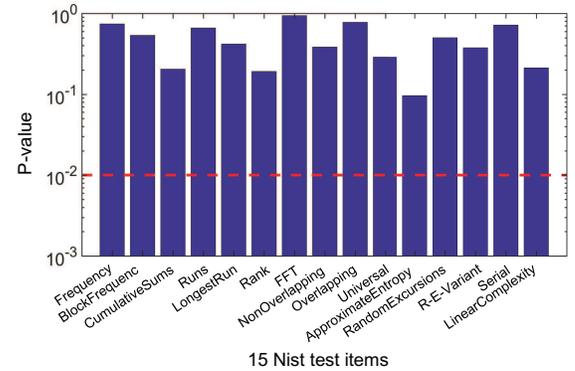}
				\caption{\label{NIST_test_result} Results of the NIST statistical test suite. The NIST standard test suites contain 15 test items. The Kolmogorov-Smirnov test is used to obtain a final p-value for the cases of multiple p-values. The test is considered successful if all final p-values satisfy 0.01$ \le $P-values$ \le $0.99.}
			\end{figure}

	To test the randomness of real-time generated random numbers, the extracted random numbers are uploaded to computer through PCIE interface for offline randomness testing. The maximum theoretical PCIE rate for transferring the extracted random numbers reaches 13.66 Gbps when the KC705 board is configured to work in 8 lanes at a 2.5 Gbps link rate mode and can effectively support our 6 Gbps data transmission.
	
	Autocorrelation tests are performed to compare the quality between raw data and extracted random numbers, as shown in FIG.\ref{auto_test}. We randomly choose $10^7$ successive raw data and $10^7$ successive  extracted random numbers. The reason for the relatively large autocorrelation existing in the raw data is that the imperfect homodyne detector without steep sideband will inevitably compromise the autocorrelation performance \cite{Shen2010Practical}, which leads to higher values of low-order autocorrelation. The autocorrelation can be significantly reduced after the Toeplitz hashing operation.
	
	We also apply the standard test suite NIST to test the obtained random numbers. 1 Gigabit binary data is read and divided into 1000 subsequences of 1 megabits. To deal with the test results, we apply Kolmogorov-Smirnov (KS) test to obtain a final p-value for the cases that each test item will obtain multiple p-values. Generally, the test is considered successful if all final p-values satisfy 0.01$ \le $P-value$ \le $0.99. The test results are shown as FIG.\ref{NIST_test_result}, which indicates that the generated random bit sequence has passed all the NIST tests.

\section*{Conclusion and Discussion}

	In this paper, we have proposed and experimentally demonstrated a high-speed, security-proved and real-time quantum random number generator by measuring vacuum fluctuation. An optimized random extraction algorithm is proposed and realized to bridge the speed gap between fast randomness generation and slow randomness extraction. The real-time rate of randomness extraction we have achieved reaches the highest speed of 12 Gbps by occupying less computing resources and the algorithm has the ability to support hundreds of Gbps randomness extraction. By assuming that the eavesdropper with complete knowledge of the classical noise, our generator has a randomness generation speed of 6.83 Gbps and this supports the generation of 6 Gbps information-theoretically provable quantum random numbers, which are output in real-time through the PCIE interface. The extraction operation eliminates potential security issues caused by classical noise. The final real-time random number output through the PCIE interface reaches 6 Gbps which is faster than the ever reported 2 Gbps\cite{Symul2011Real} scheme based on measuring vacuum fluctuation.
	
	The LO power is assumed to be constant and it is relatively stable in our experiment. But the LO power fluctuation should be investigated in further research, for the LO power can be influenced or even controlled by eavesdropper in practical issues. What’s more, the quantum randomness quantification of this work is semi-classical. Since it is a quantum random number generator, it will be interesting to apply the fully quantum analysis method\cite{Zhou2018Randomness} to quantify the extractable randomness in the future. Meanwhile, further research can be done by proposing various defense methods against hacker attacks, such as real-time min-entropy monitoring, to enhance the practical security of quantum random number generators.

	\begin{acknowledgments}
		This work was supported in part by the Key Program of National Natural Science Foundation of China under Grants No. 61531003, the National Natural Science Foundation under Grants No. 61427813, China Postdoctoral Science Foundation under Grant No. 2018M630116, and the Fund of State Key Laboratory of Information Photonics and Optical Communications.
	\end{acknowledgments}


\begin{thebibliography}{99}
		\bibitem{Ferrenberg1992} A. M. Ferrenberg, D. P. Landau, and Y. J. Wong, ``Monte Carlo simulations: Hidden errors from "good" random number generators'', Phys. Rev. Lett. \textbf{69}, 3382 (1992).
		\bibitem{Gennaro1621063} R. Gennaro, ``Randomness in cryptography'', IEEE Secur. Priv. \textbf{4}(2), 64-67  (2006).
		\bibitem{Brunner2014} N. Brunner, D. Cavalcanti, S. Pironio, V. Scarani, and S. Wehner, ``Bell nonlocality'', Rev. Mod. Phys. \textbf{86}, 419 (2014).

		\bibitem{Nisan1988Hardness} N. Nisan, ``Hardness vs. Randomness'', J. Comput. Syst. Sci. \textbf{49}, 149-167 (1994).
		\bibitem{Bouda2012Weak} J. Bouda, M. Pivoluska, M. Plesch, and C. Wilmott, ``Weak randomness seriously limits the security of quantum key distribution'', Phys. Rev. A. \textbf{86}(6), 062308 (2012).

		\bibitem{Gisin2002quantum} N. Gisin, G. Ribordy, W. Tittel, and H. Zbinden, ``Quantum cryptography'', Rev. Mod. Phys. \textbf{74}, 145 (2002).
		\bibitem{Scarani2012The} V. Scarani, H. Bechmann-Pasquinucci, N. J. Cerf, M. Du\v sek, N. L\" utkenhaus, and M. Peev, ``The security of practical quantum key distribution'', Rev. Mod. Phys. \textbf{81}, 1301 (2009).
		\bibitem{Weedbrook2012Gaussian} C. Weedbrook, S. Pirandola, R. Garc\' ia-Patr\' on, N. J. Cerf, T. C. Ralph, J. H. Shapiro, and S. Lloyd, ``Gaussian quantum information'', Rev. Mod. Phys. \textbf{84}, 621 (2012).
		\bibitem{Diamanti2016Practical}E. Diamanti, H. Lo, B. Qi, and Z. Yuan, ``Practical challenges in quantum key distribution'', npj Quantum Inf. \textbf{2}, 16025 (2016).
		
		\bibitem{Zhang2017Continuous}Y. Zhang, Z. Li, Z. Chen, C. Weedbrook, Y. Zhao, X. Wang, C. Xu, X. Zhang, Z. Wang, M. Li, X. Zhang, Z. Zheng, B. Chu, X. Gao, N. Meng, W. Cai, Z. Wang, G. Wang, S. Yu, and H. Guo, ``Continuous-variable QKD over 50km commercial fiber'', arXiv:1709.04618 (2017).
		\bibitem{Ma2016Quantum} X. Ma, X. Yuan, Z. Cao, B. Qi and Z. Zhang, ``Quantum random number generation'', npj Quantum Inf. \textbf{2}, 16021 (2016).
		\bibitem{Herrero2017quantum} M. Herrero-Collantes and J. C. Garcia-Escartin, ``Quantum Random Number Generators'', Rev. Mod. Phys. \textbf{89}, 015004 (2017).
		\bibitem{Bera2017Randomness} M. N. Bera, A. Ac\' in, M. Ku\' s, M. W. Mitchell, and M. Lewenstein, ``Randomness in Quantum Mechanics: Philosophy, Physics and Technology'', Rep. Prog. Phys. \textbf{80}, 124001 (2017).
		\bibitem{Jennewein2000A} T. Jennewein, U. Achleitner, G. Weihs, H. Weinfurter, and A. Zeilinger, ``A fast and compact quantum random number generator,'' Rev. Sci. Instrum. \textbf{71}, 1675 (2000).
		\bibitem{Andr2000Optical} A. Stefanov, N. Gisin, O. Guinnard, L. Guinnard, and H. Zbinden, ``Optical quantum random number generator'', J. Mod. Opt. \textbf{47}, 595 (2000).
		\bibitem{Ma2005Random} H. Ma, Y. Xie and L. Wu, ``Random number generation based on the time of arrival of single photons'', Appl. Opt. \textbf{44}, 7760 (2005).
		
		\bibitem{Dynes2008A} J. F. Dynes, Z. Yuan, A. W. Sharpe, and A. J. Shields, ``A high speed, postprocessing free, quantum random number generator'', Appl. Phys. Lett. \textbf{93}(3), 031109 (2008).
		\bibitem{Michael2009Photon} M. Wayne, E. Jeffrey, G. Akselrod, and P. Kwiat, ``Photon arrival time quantum random number generation'', J. Mod. Opt. \textbf{56}, 516-522 (2009).
		\bibitem{Wahl2011An} M. Wahl, M. Leifgen, M. Berlin, T. R\" ohlicke, H. Rahn and O. Benson, ``An ultrafast quantum random number generator with provably bounded output bias based on photon arrival time measurements'', Appl. Phys. Lett. \textbf{98}, 171105 (2011).
		
		\bibitem{Nie2014Practical} Y. Nie, H. Zhang, Z. Zhang, J. Wang, X. Ma, J. Zhang, and J. Pan, ``Practical and fast quantum random number generation based on photon arrival time relative to external reference'',  Appl. Phys. Lett. \textbf{104}, 051110 (2014).
		\bibitem{Wei2009Bias} W. Wei and H. Guo, ``Bias-free true random-number generator'', Opt. Lett. \textbf{34}, 1876 (2009).
		
		\bibitem{Furst2010High} H. F\" urst, H. Weier, S. Nauerth, D. G. Marangon, C. Kurtsiefer, and H. Weinfurter, "High speed optical quantum random number generation," Opt. Express \textbf{18}, 13029-13037 (2010).
		\bibitem{Ren2011Quantum} M. Ren, E Wu, Y. Liang, Y. Jian, G. Wu, and H. Zeng, ``Quantum random-number generator based on a photon-number-resolving detector'', Phys. Rev. A. \textbf{83}, 023820 (2011).
		
		\bibitem{Yan2014Multi} Q. Yan, B. Zhao Q. Liao, and N. Zhou, ``Multi-bit quantum random number generation by measuring positions of arrival photons'', Rev. Sci. Instrum. \textbf{85}, 103116 (2014).
		\bibitem{Applegate2015Efficient} M. J. Applegate, O. Thomas, J. F. Dynes, Z. Yuan, D. A. Ritchie, and A. J. Shields, ``Efficient and robust quantum random number generation by photon number detection'', Appl. Phys. Lett. \textbf{107}, 071106 (2015).
		
		\bibitem{Gabriel2010A} C. Gabriel, C. Wittmann, D. Sych, R. F. Dong, W. Mauerer,
		U. L. Andersen, C. Marquardt, and G. Leuchs, ``A generator for unique quantum random numbers based on vacuum states'', Nat. Photonics \textbf{4}, 711 (2010).
		\bibitem{Shen2010Practical} Y. Shen, L. Tian, and H. Zou, ``Practical quantum random number generator based on measuring the shot noise of vacuum states'', Phys. Rev. A \textbf{81}, 063814 (2010).
		\bibitem{Symul2011Real} T. Symul, S. M. Assad, and P. K. Lam, ``Real time demonstration of high bitrate quantum random number generation with coherent laser light'', Appl. Phys. Lett. \textbf{98}, 231103 (2011).
		\bibitem{Haw2015Maximization} J. Y. Haw, S. M. Assad, A. M. Lance, N. H. Y. Ng, V. Sharma, P. K. Lam, and T. Symul, ``Maximization of Extractable Randomness in a Quantum Random Number Generator'', Phys. Rev. Appl. \textbf{3}, 054004 (2015).
		
		\bibitem{Bingjie2017High} B. Xu, Z. Li, J. Yang, S. Wei, Q. Su, W. Huang, Y. Zhang, and H. Guo, ``High Speed Continuous Variable Source-Independent Quantum Random Number Generation'', Quantum Sci. Technol. \textbf{4}, 025013 (2019).
		
		\bibitem{Zhou2017Practical}	Q. Zhou, R. Valivarthi, C. John, and W. Tittel,  ``Practical quantum random number generator based on sampling vacuum fluctuations'', arXiv:1703.00559 (2017).
		
		\bibitem{Santamato2017An} F. Raffaelli, G. Ferranti, D. H Mahler, P. Sibson, J. E. Kennard, A. Santamato, G. Sinclair, D. Bonneau, M. G Thompson, and J. C F Matthews, ``A homodyne detector integrated onto a photonic chip for measuring quantum states and generating random numbers'', Quantum Sci. Technol. \textbf{3}, 025003 (2018).
		\bibitem{Qi2010Highspeed} B. Qi, Y. Chi, H. Lo, and L. Qian, ``High-speed quantum random number generation by measuring phase noise of a single-mode laser'', Opt. Lett. \textbf{35}, 312-314 (2010).
		\bibitem{Guo2010Truly}H. Guo, W. Tang, Y. Liu, and W. Wei, ``Truly random number generation based on measurement of phase noise of a laser'', Phys. Rev. E \textbf{81}, 051137 (2010).
		\bibitem{Xu2012Ultrafast} F. Xu, B. Qi, X. Ma, H. Xu, H. Zheng, and H. Lo ``Ultrafast quantum random number generation based on quantum phase fluctuations'', Opt. Express \textbf{20}, 12366 (2012).
		\bibitem{Abellan2014Ultra} C. Abell\' an, W. Amaya, M. Jofre, M. Curty, A. Ac\' in, J. Capmany, V. Pruneri, and M. W. Mitchell, "Ultra-fast quantum randomness generation by accelerated phase diffusion in a pulsed laser diode," Opt. Express \textbf{22}, 1645-1654 (2014).
		
		\bibitem{Nie2015The} Y. Nie, L. Huang, Y. Liu, F. Payne, J. Zhang, and J. Pan, ``The generation of 68 Gbps quantum random number by measuring laser phase fluctuations'', Rev. Sci. Instrum. \textbf{86}, 063105 (2015).
		\bibitem{Yang2016A} J. Yang, J. Liu, Q. Su, Z. Li, F. Fan, B. Xu, and H. Guo, ``5.4 Gbps real time quantum random number generator with compact implementation'', Opt. Express \textbf{24}, 27475 (2016).
		\bibitem{Zhang2016Note} X. Zhang, Y. Nie, H. Zhou, H. Liang, X. Ma, J. Zhang, and J. Pan, ``Fully integrated 3.2 Gbps quantum random number generator with real-time extraction'', Rev. Sci. Instrum. \textbf{87}, 076102 (2016).
		
		\bibitem{Liu2017117} J. Liu, J. Yang, Q. Su, Z. Li, F. Fan, B. Xu, and H. Guo, ``117 Gbits/s quantum random number generation with simple structure'', IEEE Photon. Technol. Lett. \textbf{29}, 1109 (2017).
		
		\bibitem{Williams2010Fast} C. R. S. Williams, J. C. Salevan, X. Li, R. Roy, and T. E. Murphy, ``Fast physical random number generator using amplified spontaneous emission'', Opt. Express \textbf{18}, 23584 (2010).		
		\bibitem{Li2011Scalable} X. Li, A. B. Cohen, T. E. Murphy, and R. Roy, ``Scalable parallel physical random number generator based on a superluminescent LED'', Opt. Lett. \textbf{36}, 1020 (2011).
		\bibitem{Wei2012High} W. Wei, G. Xie, A. Dang, and H. Guo, ``High-speed and bias-free optical random number generator'', IEEE Photon. Technol. Lett. \textbf{24}, 437 (2012).
		
		\bibitem{Liu2013Implementation} Y. Liu, M. Zhu, B. Luo, J. Zhang, and H. Guo, ``Implementation of 1.6 Tbs-1 truly random number generation based on a super-luminescent emitting diode'', Laser Phys. Lett. \textbf{10}, 045001 (2013).
		
		\bibitem{Qi2017True} B. Qi, ``True randomness from an incoherent source,'' Rev. Sci. Instrum. \textbf{88}, 113101 (2017).
		
		\bibitem{Martin2015Quantum} A. Martin, B. Sanguinetti, C. C. W. Lim, R. Houlmann, and H. Zbinden, ``Quantum random number generation for 1.25-GHz quantum key distribution systems'', J. Lightwave Technol. \textbf{33}, 2855 (2015).
		
		
		\bibitem{Impagliazzo1989Pseudo} R. Impagliazzo, L. A. Levin , and M. Luby, ``Pseudo-random generation from one-way functions'', in \textit{Proceedings of the 21st Annual ACM Symposium Theory of Computing (STOC)} (ACM, 1989), pp.12–24.
		\bibitem{Mansour1990The} Y. Mansour, N. Nisan, and P. Tiwari, ``The computational complexity of universal hashing'', in \textit{Proceedings of the 22nd Annual ACM Symposium Theory Computing (STOC)} (ACM, 1990), pp. 235–243.
		\bibitem{Ma2012Postprocessing} X. Ma, F. Xu, H. Xu, X. Tan,	B. Qi, and H. Lo, ``Postprocessing for quantum random-number generators: Entropy evaluation and randomness extraction'', Phys. Rev. A \textbf{87}, 062327 (2013).
		\bibitem{Wang2018High} X. Wang, Y. Zhang, S. Yu, and H. Guo, ``High-Speed Implementation of Length-Compatible Privacy Amplification in Continuous-Variable Quantum Key Distribution'', IEEE Photonics Journal, \textbf{10}(3), 7600309 (2018).
		\bibitem{ZHANG20181.2G} X. Zhang, Y. Zhang, Z. Li, S. Yu and H. Guo, ``1.2-GHz Balanced Homodyne Detector for Continuous-Variable Quantum Information Technology'', IEEE Photonics Journal, \textbf{10}(5), 6803810 (2018).
		\bibitem{Zhou2018Randomness} H. Zhou, P. Zeng, M. Razavi, and X. Ma, ``Randomness quantification of coherent detection'', Phys. Rev. A \textbf{98}, 042321 (2018).

		
		
		
		
	\end{thebibliography}
\end{document}